\documentclass[10pt]{blois07}
\usepackage{graphicx}
\usepackage{amsmath}
\usepackage{cite,./mcite}

\begin{document}
\title{Connections between high energy QCD and statistical physics}
\author{St\'ephane Munier}
\institute{Centre de physique th\'eorique, \'Ecole polytechnique, CNRS,
91128 Palaiseau, France
}
\maketitle
\begin{abstract}
It has been proposed that the energy
evolution of QCD amplitudes in the high-energy regime
falls in the universality class of reaction-diffusion processes.
We review the arguments for this correspondence, and we explain how it
enables one to compute analytically asymptotic features of QCD amplitudes.
\end{abstract}


The high-energy regime of QCD has been intensively studied in deep-inelastic
$e$-$p$ scattering at HERA, 
in heavy-ion collisions at RHIC,
and will be probed in proton and heavy-ion scattering at LHC.
It is rich of interesting theoretical 
structures: Links have been found or conjectured
with conformal field theory, 
string theory,
and more recently, with a class of models 
known in statistical physics.

When hadrons scatter at very high energy,
the color fields that are generated at the interaction point
have a large strength.
Perturbative methods qualify as soon
as there is a large transverse momentum scale in the event:
This property enables one, under certain conditions, 
to derive QCD evolution equations,
in the form of partial differential equations
(which can also be stochastic in particular formulations such as the dipole
model \cite{Mueller:1993rr}).
On the other hand, strong fields cause the parton densities to
saturate, which makes this evolution nonlinear.

Similar looking stochastic nonlinear partial differential equations
also appear in problems of apparently different physical origin,
such as reaction-diffusion, or population evolution.
The goal of this short review is to explain that these similarities
are not casual, and that once understood, they can help the
derivation of new results for QCD cross sections.
We refer the reader to the original papers 
\cite{Munier:2003vc,*Munier:2003sj,*Munier:2004xu,Iancu:2004es,*Munier:2005re} 
for the details,
and to Ref.~\cite{Munier:2006zf} for
a more extensive review.

In the following, we will consider the scattering of two hadrons, 
and we will aim
at computing their cross section
at very high energies. 
Their relative rapidity is denoted by $Y$.
Since our discussion
will rely on resummed perturbative QCD, we think of
these hadrons as being small objects, 
such as color dipoles found for example as
fluctuations of highly virtual photons.
We will always be discussing a definite region of impact
parameter.

\section{High energy QCD and reaction-diffusion}

Cross sections
are measured by counting the number of events that are registered 
in a detector within a given interval of time.
Each single event results from an interaction between
the scattering hadrons {\it realized as definite quantum states}, that is,
as particular Fock states. 
Let us go to the frame in which one hadron is almost at rest,
while the other one 
carries most of the kinetic energy, and thus
develops a highly-occupied Fock state.
As long as saturation effects 
are negligible (i.e. far from
the unitarity limit in which 
the hadrons appear black to each other), 
the probability of interaction
is proportional to the number of partons in the fast hadron whose
transverse momenta $k$ match the typical momentum scale of
the slow hadron.
Let us imagine that $k$ is tunable (it is the case 
when the slow object is a virtual photon),
and that one could actually measure the interaction probability
of the slow hadron
with {\em a particular Fock state} of the fast hadron.
(In practice, it would require
the replication of quantum Fock states, which is impossible).
We  call $T(k)$ this interaction probability (more precisely, it is
the forward elastic scattering amplitude at a fixed impact parameter).
$T$ is an unphysical quantity, but it will be important
to understand it theoretically.
The physical amplitude $A(Y,k)$ is just
the average of $T(k)$ over all possible Fock state realizations
at the considered rapidity $Y$:
$A=\langle T\rangle$, that is, $A$ is the average over
all events that may occur, appropriately weighted by 
their probabilities at a given rapidity.

Standard quantum field theory calculations, 
based on the evaluation
of Feynman diagrams, would directly lead to the expression 
of~$A$.
However,
it turns out that such calculations are extremely hard.
Instead, understanding first the main analytical 
features of the scattering probability $T$ off a single
typical Fock state of the fast hadron and then
averaging over events is a simpler approach, that has been
successful in leading to analytical expressions for
the asymptotics of the
scattering cross sections.

What is precisely known about $T$ is its evolution with rapidity
(or energy), at least in the regime in which $T\ll 1$, that
is, away from the unitarity limit.
When one increases infinitesimally
 the rapidity of a hadron that has say $n$ partons
in its current Fock state,
there is a transition rate to a $n+1$, $n+2$...--parton Fock state 
that is computable in perturbative QCD.
It may be extracted from the BFKL equation.
A direct formulation is the color dipole model \cite{Mueller:1993rr}, 
in which this transition
probability is explicitely computed.
When $T\ll 1$, $T$ is a linear function of the number of partons.
Roughly, it reads $T(k)\sim \alpha_s^2 n(k)$, where $n(k)$ is
the number of gluons
that have a transverse momentum of the order of $k$.

This transition to higher Fock states may be captured by a linear
stochastic equation, of the form
\begin{equation}
\partial_{\bar\alpha Y}T=\chi(-\partial_{\ln k^2})T+\alpha_s\sqrt{T}\,\nu,
\end{equation}
where $\bar\alpha=\alpha_s N_c/\pi$.
$\nu$ is an appropriate stochastic variable 
that has zero mean, and variations of order unity when $\bar\alpha Y$
is increased by one.
$\chi(-\partial_{\ln k^2})$ is the usual BFKL kernel.
It describes the branching diffusion of partons, at least in the
regime of very high energy in which we are interested in. This means
that, when acted on $T$, it roughly behaves like 
a diffusion term
$\partial_{\ln k^2}^2 T$ supplemented by a growth term $T$
(with appropriate coefficients).
The noise term is a consequence of discreteness: It implements the 
fact that we are considering the evolution of one single Fock state,
that contains a definite (discrete) number of partons.

\begin{figure}
\begin{center}
\includegraphics[angle=-90,width=0.9\textwidth]{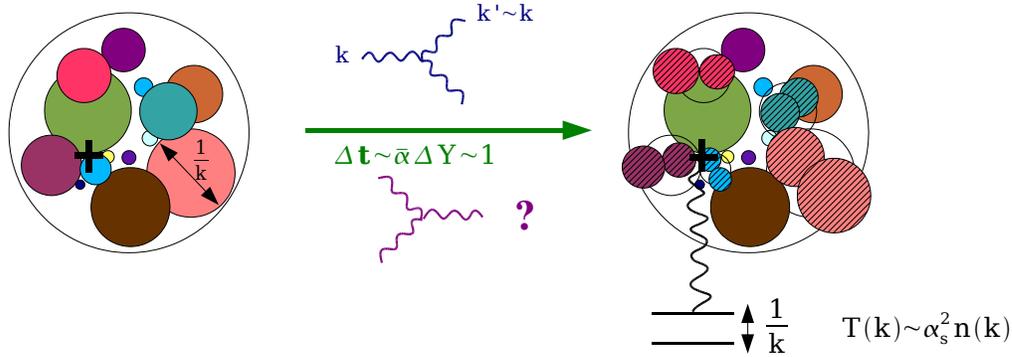}
\end{center}
\caption{\label{sketch}
{\em Sketch of the gluonic content (the gluons are represented by disks)
of the hadron in the transverse plane.}
In a little boost, each gluon can split, or nonlinear
effects like recombination can take place.
At the time of the interaction, the system is probed by a
dipole of size $1/k$, which is sensitive to the number of gluons
of similar size at a given impact parameter.
}
\end{figure}

On the other hand, when $T$ becomes of the order of $1$, {\em saturation
effects} have to enter in order
to tame the growth of the number of partons,
for unitarity to be preserved. From the work 
of Balitsky and Kovchegov (BK), we know
that in the mean-field limit in which the noise can be neglected (that is, 
when $A=T$; This is realized when one of the interacting objects
is a very large nucleus), 
the evolution equation for $T$ reads
\begin{equation}
\partial_{\bar\alpha Y}T=\chi(-\partial_{\ln k^2})T-T^2.
\label{bk}
\end{equation}
Hence we shall propose that the full evolution 
be described by the following stochastic equation:
\begin{equation}
\partial_{\bar\alpha Y}T=\chi(-\partial_{\ln k^2})T-T^2+\alpha_s\sqrt{T}\,\nu.
\label{stoch}
\end{equation}
This equation is in the universality class of the stochastic version of the
Fisher and Kolmogorov, Petrovsky, Piscounov (F-KPP) equation.
(The latter would in fact be obtained
by replacing $\chi(-\partial_{\ln k^2})T$ by $\partial_{\ln k^2}T+T$
and $\sqrt{T}$ by $\sqrt{T(1-T)}$; for a review and references, 
see~\cite{panja-2004-393}.)
A schematic picture of the evolution is presented 
in Fig.~\ref{sketch}.

So far, no one has succeeded in formulating rigorously scattering in
QCD in the form of a stochastic equation such as~(\ref{stoch}),
in particular for the way how saturation occurs
is not yet fully understood.
Is it gluon recombination, as was advocated in the early
papers on saturation \cite{Gribov:1981ac,*Mueller:1985wy}, 
or some more subtle process?
So the best one can do at this stage 
is to set the noise term in such a way that 
the evolution of the hadron Fock state by gluon splittings
is exactly reproduced
away from the unitarity limit, where
the nonlinear term may be neglected.
A practical implementation of this process would be, 
for example, Salam's 
Monte-Carlo code of the dipole model \cite{Salam:1996nb}
modified by the addition of a suitable
saturation condition which
makes sure that $T$ (written in coordinate space) keeps always less 
than 1 and thus
that unitarity is preserved.
In this procedure, 
the expression for the noise is unambiguously fixed (it
results from the splitting probability of the dipoles), and the BFKL
limit is exactly taken into account.
Such a procedure was suggested in Ref.~\cite{Munier:2006um}, but
has not been implemented so far for its technical awkwardness.
Some other paths were followed:
One may alternatively take $\nu$ to be a Gaussian white noise\footnote{%
However, in this case, $T$ should not exactly be the amplitude, but
rather a kind of ``dual amplitude'' -- see for example Ref.~\cite{gardiner}.} 
\cite{Iancu:2004iy}. This
simple choice enables one to
apply the Ito stochastic calculus, and to draw a link
with equations established within QCD such as the B-JIMWLK equations
(For a review and references, see Ref.~\cite{Weigert:2005us}, and A.~Shoshi's talk at
this conference \cite{shoshi-2007}.) 
One may also think of the whole
process as a reaction-diffusion process, 
as we will implicitely do in the next section.

All this may look quite arbitrary: We have merely merged two
known limits into a single equation, without much further 
justification.
How can one be sure that $A$ obtained from averaging
realizations of Eq.~(\ref{stoch}) 
looks like the solution of a genuine QCD equation?
Although it might sound strange a priori, 
the solution to the ``arbitrary'' equation
that we have written down
is very likely to contain the exact asymptotics of QCD.
This fact is actually related to the universality of the solutions
to such evolution equations.
The statement is the following: For a large class of processes, i.e.
for a number of stochastic functions $\nu$
and for a variety of forms of the nonlinearities,
the asymptotics of the statistics of $T$ (that is, the physical observables
$A\equiv\langle T\rangle$, $\langle T^2\rangle$...) for small $\alpha_s$ and
large $\bar\alpha Y$ are identical.
This is not a theorem, but a conjecture based
on a general understanding of how noisy traveling waves propagate.
(The propagation mechanism is described in the next section.)
The whole 
point is that the details of the evolution equation
do not matter for extracting the asymptotics
of the QCD amplitudes.

Let us describe a typical process whose evolution is
in the universality class described
by the F-KPP equation:
reaction-diffusion. This process
involves particles on a lattice indexed by some variable $x$, 
that evolve by a set of rules of the following form: 
As time is increased, each particle
has a probability either to jump to a nearby site, or to
become two particles, or to recombine with another particle
on the same lattice site. The balance between creation
and recombination of particles determines the equilibrium number
of particles on each site $N$. 
After a large evolution time, the number of particles 
on a given site oscillates
about $N$ (with an amplitude of the order of 
the typical statistical fluctuations $\sqrt{N}$).
It is the number of particles per site normalized to $N$
that obeys an equation in the universality class of the F-KPP equation.

At this point, we may establish 
a simple dictionary between reaction-diffusion
and QCD.
Time is the evolution variable, so is rapidity: 
$t\leftrightarrow\bar\alpha Y$. The variable in which diffusion
takes place is $x\leftrightarrow\ln k^2$. 
The equilibrium number of particles is $N\leftrightarrow 1/\alpha_s^2$.
(In QCD, it is fixed by the unitarity condition\footnote{%
This condition actually holds in coordinate space (when $T$ is a function of
transverse sizes). In momentum space, the growth of $T$
with energy is also tamed as soon as 
the point $T=1$ has been crossed,
although $T(k)$ can take arbitrarily large values.
This does not change the conclusions that we shall draw later: 
The only important feature of the evolution is that $T$ changes 
behavior in the saturation region. One can see how it goes precisely
in QCD e.g. in the numerical
simulations presented in Ref.~\cite{Enberg:2005cb}.
}  $T\leq 1$.)


\section{Statistical methods and application to QCD}

\begin{figure}
\begin{center}
\includegraphics[angle=-90,width=0.5\textwidth]{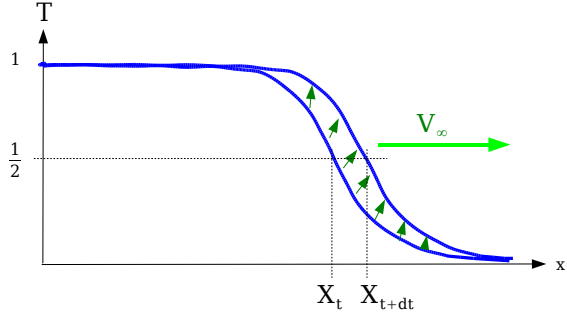}
\end{center}
\caption{\label{mean0}
{\em Deterministic F-KPP front and its evolution with time.}
The arrows show where branching diffusion takes place to drive the
motion towards larger values of $x$.
}
\end{figure}

In a first step, we ignore the stochastic term, that is, we 
address the
simple BK equation~(\ref{bk}), in order to gain intuition
on the form of the solutions. A given localized initial condition 
($T\sim \alpha_s^2$ in
a region of order 1 around some initial scale $\ln k_0^2$: 
This would be the physical initial condition) will
spread and grow under the action of the kernel $\chi(-\partial_{\ln k^2})$,
which, as we wrote before, amounts to a branching diffusion.
But as soon as $T$ becomes of the order of 1, the nonlinear term
enters to compensate the growth, making $T$ saturate.
Then further evolution necessarily has the form of two symmetric 
traveling waves, since the system can only escape to the right and to
the left. Let us focus on the rightmoving wave, that travels
towards larger values of $\ln k^2$. This wave front is represented
schematically in Fig.~\ref{mean0}.
It turns out that the shape of this wave in its large-$\ln k^2$ 
tail is exponential, with a slope that is completely fixed
by the linear kernel:
\begin{equation}
T\sim e^{-\gamma_0\ln k^2},\ \ \ 
\text{$\gamma_0$ being determined by}\ \ \ 
\frac{\chi(\gamma_0)}{\gamma_0}=\chi^\prime(\gamma_0).
\label{expo}
\end{equation}
Since the wave front keeps its shape, it makes sense to
characterize its motion by a single velocity $V_\infty$.
The latter is also completely determined by the kernel $\chi$. It reads
\begin{equation}
V_\infty=\frac{\chi(\gamma_0)}{\gamma_0}.
\end{equation}
In QCD, the position $X_t$ of the wave is called the saturation scale
$Q_s(Y)$. It characterizes the momentum below which nonlinear saturation
effects (gluon recombination, multiple scattering...) become important.
The velocity $V_\infty$ defined above is simply the derivative of 
$\ln Q_s^2(Y)$ with respect to $Y$. (Recall that the $x$-variable is $\ln k^2$.)

Now that we have understood the deterministic limit,
we may try to put back the noise.
We do not know how to attack the problem directly.
Instead, we shall stick to a physical approach, and view the evolution 
equation~(\ref{stoch}) as describing a reaction-diffusion process.
In this framework, we recall that the origin 
of the noise was the discreteness
of the number of particles on each site.
Discreteness means in particular that the number of particles
$n(k)$ cannot be a fraction of an integer. Consequently,
coming back to the simple-minded relationship\footnote{%
Again, this is not literally true: $T(k)$ is actually continuous,
but the tails (below $T=\alpha_s^2$)
are decaying exponentially with a characteristic length of 
one unit in the variable
$\ln k^2$. This is steep enough for all our arguments to apply as
if $T(k)$ itself were discrete.
} $T(k)\sim\alpha_s^2 n(k)$,
it means that $T$ is either 0 are larger than $\alpha_s^2$.
Brunet and Derrida \cite{brunet-1997-57}
proposed to replace the full stochastic equation
by a deterministic one that takes into account this basic
effect of discreteness,
which can easily be done by not allowing any growth
when $T<\alpha_s^2$. (It amounts to cutting off the tail of the front; 
to do this in practice, one may for example replace $\chi$ by
a modified kernel obtained by subtracting
its growth term in the region in which $T<\alpha_s^2$. 
Note that there is
no unique prescription.)
The solution to this modified equation is again a traveling
wave, that exhibits the same overall exponential decay as given
by Eq.~(\ref{expo})
(except for an uninteresting additional prefactor).
Its velocity now reads\footnote{%
This result had already been obtained by 
Mueller and Shoshi \cite{Mueller:2004sea}.
Actually, 
the understanding of high-energy scattering as
a peculiar reaction-diffusion process emerged
from a reinterpretation of their work, in the light of the Brunet-Derrida
analytical treatment of traveling waves with a 
cutoff~\cite{brunet-1997-57}.
}
\begin{equation}
V_\text{BD}
=\frac{\chi(\gamma_0)}{\gamma_0}-\frac{\pi^2\gamma_0\chi^{\prime\prime}(\gamma_0)}
{2\ln^2(1/\alpha_s^2)}.
\end{equation}
It is thus less than the velocity of the front in the limit of an
infinite number of particles (obtained by letting $\alpha_s$ go to 0), 
which had
to be expected: Indeed, taking into account discreteness amounts
to removing some ``matter'' from the front, which logically 
slows down its motion.

\begin{figure}
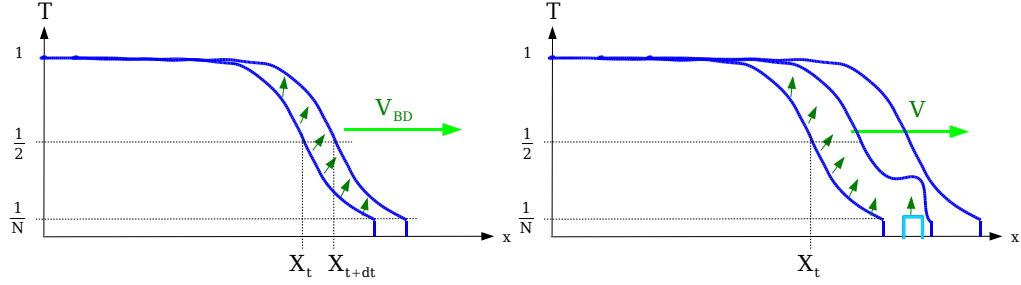

\begin{center}
\includegraphics[angle=-90,width=0.45\textwidth]{mean1.epsi}
\includegraphics[angle=-90,width=0.45\textwidth]{mean2.epsi}
\end{center}
\caption{\label{mean1}{\em Traveling waves solution of the F-KPP equation 
with a cutoff that simulates discreteness.} In the left sketch,
no stochasticity is taken into account while in the right one, particles may
randomly be sent ahead of the deterministic front. Their further time/rapidity
evolution is also represented.}
\end{figure}

However, a deterministic solution can only reproduce approximately the
realizations of a stochastic evolution.
We can incorporate stochasticity back into the picture
\cite{Brunet:2005bz}
by noting once again that the noise is only important
in the forward tail of the front, where the number of particles 
is low on the average.
From numerical simulations of simple reaction-diffusion models, 
we observed the following behavior: 
Most of the time, the motion of the front is
almost deterministic, with a velocity given by the solution 
to the cutoff deterministic equation.
From time to time, rarely, a large fluctuation causes a transitory 
acceleration of the front. This fluctuation consists in one or a few
particles being sent 
far ahead of the deterministic tip of the front, which then evolve
into a new front that later gets absorbed by the deterministic front.
This behavior is represented in Fig.~\ref{mean1}.
We conjectured a probability distribution for these fluctuations,
as well as the effect that they have on the position of the front
after relaxation.

With these elements, we were able to
deduce the full statistics of the saturation scale, 
that is to say not only the
mean position (or velocity) of the front,
\begin{equation}
V=\frac{\langle \ln Q_s^2\rangle}{\bar\alpha Y}
=\frac{\chi(\gamma_0)}{\gamma_0}
-\frac{\pi^2\gamma_0\chi^{\prime\prime}(\gamma_0)}
{2\ln^2(1/\alpha_s^2)}+\pi^2\gamma_0^2\chi^{\prime\prime}(\gamma_0)
\frac{3\ln\ln(1/\alpha_s^2)}{\gamma_0\ln^3(1/\alpha_s^2)},
\label{velocity}
\end{equation}
 but also all its cumulants:
\begin{equation}
{\langle \ln^n Q_s^2\rangle_\text{cumulant}}=
\pi^2\gamma_0^2\chi^{\prime\prime}(\gamma_0)
\frac{n!\zeta(n)}{\gamma_0^n}
\frac{\bar\alpha Y}{\ln^3(1/\alpha_s^2)},
\label{cumulants}
\end{equation}
when $n\geq 2$.

Now we recall that the physical amplitude is obtained by 
averaging $T$
over all possible realizations.
Given that the fall off of the large-$\ln k^2$ tail of 
each single event is exponential,
it is not difficult to get the scaling of the scattering amplitude
with the help of Eq.~(\ref{cumulants}):
\begin{equation}
A(Y,k)=A\left(
\frac{\ln{k^2}-\langle\ln{ Q_s ^2(Y)}\rangle}
{\sqrt{\frac{\bar\alpha Y}{\ln^3(1/\alpha_s^2)}}}
\right),
\label{xsec}
\end{equation}
where $\langle\ln{ Q_s ^2(Y)}\rangle$ is given by Eq.~(\ref{velocity}).
This is the main analytical result for QCD that comes out
of the statistical approach.
Note that other results can be extracted on the statistics
of the branchings of the gluons in the course of the evolution,
but we cannot see a possible phenomenological application.

\begin{figure}
\begin{center}
\includegraphics[angle=-90,width=0.8\textwidth]{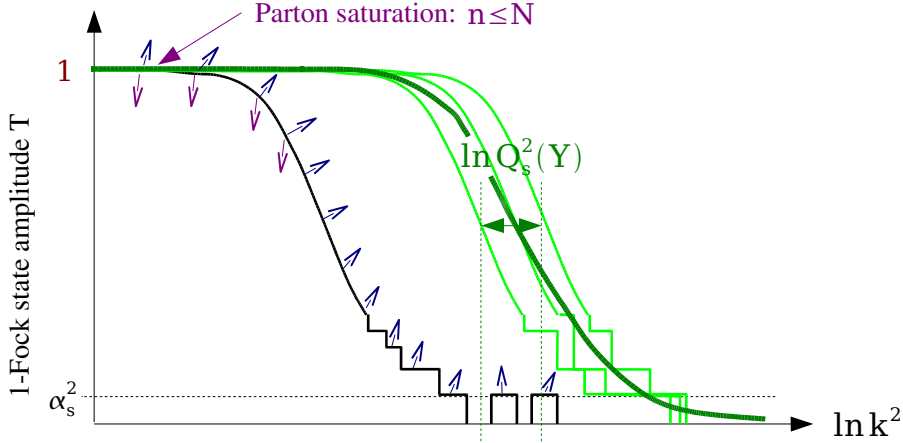}
\end{center}
\caption{\label{stocha}{\em Time/rapidity 
evolution of a noisy traveling wave.}
The noise is essentially concentrated at the tip of the front, where
the occupation numbers are low.
After some time (rapidity), the wave has moved to the right (3 realizations
are shown with thin lines),
roughly keeping its shape.
However, stochasticity manifests itself macroscopically by
inducing a dispersion in the positions of the fronts between
different realizations.
Since the physical amplitude $A$ is the average of all realizations,
its very shape is influenced by the noise.
($A$ is represented by the thick line.)
}
\end{figure}

The emerging overall picture 
of front propagation is shown in Fig.~\ref{stocha}.

\section{Prospects}

Clearly, the statistical interpretation of scattering processes
has proved useful since it has led to both a new understanding
and new asymptotical results for high energy QCD.
Of course, it relies on a few conjectures that will eventually have
to be proved in a more formal way, but we feel that we have
so far provided robust physical arguments.

It has to be acknowledged that our new analytical results
are not relevant for phenomenology yet, since they make sense
for $\ln (1/\alpha_s^2)\gg 1$ only, 
which requires values of $\alpha_s$ so small
that, of course, they are far beyond
the experimentally attainable range. 
A number of authors have however taken seriously the 
extrapolation of these results to realistic values of $\alpha_s$ 
and have produced predictions,
see e.g. 
Ref.~\cite{Hatta:2006hs,*Kozlov:2006qw,*Kozlov:2007wm}.
On the other hand, numerics could give results valid for 
$\alpha_s<0.1$ (optimistically),
that is, not far from the phenomenological domain. 
(This point is discussed in
Ref.~\cite{Munier:2006um}).

At this point, we have been able to extract
properties that QCD shares with simple statistical models. 
We could claim that this was a correct procedure 
because asymptotic properties of the solutions
do not depend on the details of how saturation occurs.
So in some sense, we have done the
``easy'' part of the work. However, to go closer to 
phenomenology, one would need to understand more deeply
the details of saturation, which probably constrain the
form of the noise $\nu$ in Eq.~(\ref{stoch}).
Investigations of some possible models have been conducted,
sometimes leading to pecularities 
in the interpretation,
such as negative transition
rates \cite{Iancu:2005dx,*Bondarenko:2006rh}.
Building a complete picture, valid beyond asymptotics, 
remains a challenging open question, for which
a further breakthrough may be needed.

Finally, our approach to the propagation of noisy traveling waves
is not based on a field theory formulation,
but is an event-by-event analysis of the shape of realizations,
using methods more familiar to statistical physicists
than to particle physicists.
Being able to recover results such as 
Eqs.~(\ref{velocity}),~(\ref{cumulants}),~(\ref{xsec}) 
within field theory,
starting e.g. from an effective Lagrangian whose building blocks
are Reggeon fields,
would be a very interesting achievement.
Some progress has been made recently, see e.g. Ref.~\cite{Levin:2007wc}.

\begin{footnotesize}

\providecommand{\etal}{et al.\xspace}
\providecommand{\href}[2]{#2}
\providecommand{\coll}{Coll.}
\catcode`\@=11
\def\@bibitem#1{%
\ifmc@bstsupport
  \mc@iftail{#1}%
    {;\newline\ignorespaces}%
    {\ifmc@first\else.\fi\orig@bibitem{#1}}
  \mc@firstfalse
\else
  \mc@iftail{#1}%
    {\ignorespaces}%
    {\orig@bibitem{#1}}%
\fi}%
\catcode`\@=12


\begin{mcbibliography}{10}

\bibitem{Mueller:1993rr}
A.~H. Mueller,
\newblock Nucl. Phys.{} {\bf B415},~373~(1994)\relax
\relax
\bibitem{Munier:2003vc}
S.~Munier and R.~Peschanski,
\newblock Phys. Rev. Lett.{} {\bf 91},~232001~(2003)\relax
\relax
\bibitem{Munier:2003sj}
S.~Munier and R.~Peschanski,
\newblock Phys. Rev.{} {\bf D69},~034008~(2004)\relax
\relax
\bibitem{Munier:2004xu}
S.~Munier and R.~Peschanski,
\newblock Phys. Rev.{} {\bf D70},~077503~(2004)\relax
\relax
\bibitem{Iancu:2004es}
E.~Iancu, A.~H. Mueller, and S.~Munier,
\newblock Phys. Lett.{} {\bf B606},~342~(2005)\relax
\relax
\bibitem{Munier:2005re}
S.~Munier,
\newblock Nucl. Phys.{} {\bf A755},~622~(2005)\relax
\relax
\bibitem{Munier:2006zf}
S.~Munier,
\newblock Acta Phys. Polon.{} {\bf B37},~3451~(2006)\relax
\relax
\bibitem{panja-2004-393}
D.~Panja,
\newblock Physics Reports{} {\bf 393},~87\relax
\relax
\bibitem{Gribov:1981ac}
L.~V. Gribov, E.~M. Levin, and M.~G. Ryskin,
\newblock Nucl. Phys.{} {\bf B188},~555~(1981)\relax
\relax
\bibitem{Mueller:1985wy}
A.~H. Mueller and J.-w. Qiu,
\newblock Nucl. Phys.{} {\bf B268},~427~(1986)\relax
\relax
\bibitem{Salam:1996nb}
G.~P. Salam,
\newblock Comput. Phys. Commun.{} {\bf 105},~62~(1997)\relax
\relax
\bibitem{Munier:2006um}
S.~Munier,
\newblock Phys. Rev.{} {\bf D75},~034009~(2007)\relax
\relax
\bibitem{gardiner}
C.~G. Gardiner,
\newblock {\em Handbook of Stochastic Methods}.
\newblock Springer, 2004\relax
\relax
\bibitem{Iancu:2004iy}
E.~Iancu and D.~N. Triantafyllopoulos,
\newblock Nucl. Phys.{} {\bf A756},~419~(2005)\relax
\relax
\bibitem{Weigert:2005us}
H.~Weigert,
\newblock Prog. Part. Nucl. Phys.{} {\bf 55},~461~(2005)\relax
\relax
\bibitem{shoshi-2007}
A.~I. Shoshi~(2007).
\newblock \href{http://www.arXiv.org/abs/arXiv:0708.4322 [hep-ph]}{{\tt
  arXiv:0708.4322 [hep-ph]}}\relax
\relax
\bibitem{Enberg:2005cb}
R.~Enberg, K.~J. Golec-Biernat, and S.~Munier,
\newblock Phys. Rev.{} {\bf D72},~074021~(2005)\relax
\relax
\bibitem{Mueller:2004sea}
A.~H. Mueller and A.~I. Shoshi,
\newblock Nucl. Phys.{} {\bf B692},~175~(2004)\relax
\relax
\bibitem{brunet-1997-57}
E.~Brunet and B.~Derrida,
\newblock Physical Review E{} {\bf 57},~2597\relax
\relax
\bibitem{Brunet:2005bz}
E.~Brunet, B.~Derrida, A.~H. Mueller, and S.~Munier,
\newblock Phys. Rev.{} {\bf E73},~056126~(2006)\relax
\relax
\bibitem{Hatta:2006hs}
Y.~Hatta, E.~Iancu, C.~Marquet, G.~Soyez, and D.~N. Triantafyllopoulos,
\newblock Nucl. Phys.{} {\bf A773},~95~(2006)\relax
\relax
\bibitem{Kozlov:2006qw}
M.~Kozlov, A.~I. Shoshi, and B.-W. Xiao,
\newblock Nucl. Phys.{} {\bf A792},~170~(2007)\relax
\relax
\bibitem{Kozlov:2007wm}
M.~Kozlov, A.~Shoshi, and W.~Xiang~(2007).
\newblock \href{http://www.arXiv.org/abs/arXiv:0707.4142 [hep-ph]}{{\tt
  arXiv:0707.4142 [hep-ph]}}\relax
\relax
\bibitem{Iancu:2005dx}
E.~Iancu, G.~Soyez, and D.~N. Triantafyllopoulos,
\newblock Nucl. Phys.{} {\bf A768},~194~(2006)\relax
\relax
\bibitem{Bondarenko:2006rh}
S.~Bondarenko, L.~Motyka, A.~H. Mueller, A.~I. Shoshi, and B.~W. Xiao,
\newblock Eur. Phys. J.{} {\bf C50},~593~(2007)\relax
\relax
\bibitem{Levin:2007wc}
E.~Levin, J.~Miller, and A.~Prygarin~(2007).
\newblock \href{http://www.arXiv.org/abs/arXiv:0706.2944 [hep-ph]}{{\tt
  arXiv:0706.2944 [hep-ph]}}\relax
\relax
\end{mcbibliography}

\end{footnotesize}
\end{document}